\documentclass[twocolumn]{aastex631}

\shorttitle{No Jupiter Analog in the Eta Cassiopeiae System}
\shortauthors{Stephen R. Kane}

\begin{document}

\title{No Giant Planets in the Eta Cassiopeiae System: Dynamical
  Implications of a Wide Binary Companion}

\author[0000-0002-7084-0529]{Stephen R. Kane}
\affiliation{Department of Earth and Planetary Sciences, University of
  California, Riverside, CA 92521, USA}
\email{skane@ucr.edu}

\author[0000-0002-4860-7667]{Zhexing Li}
\affiliation{Department of Earth and Planetary Sciences, University of
  California, Riverside, CA 92521, USA}

\author[0000-0002-0139-4756]{Michelle L. Hill}
\affiliation{Department of Earth and Planetary Sciences, University of
  California, Riverside, CA 92521, USA}

\author[0000-0001-5592-6220]{Skylar D'Angiolillo }
\affiliation{Department of Earth and Planetary Sciences, University of
  California, Riverside, CA 92521, USA}

\author[0000-0003-3504-5316]{Benjamin J. Fulton}
\affiliation{Department of Astronomy, California Institute of
  Technology, Pasadena, CA 91125, USA}

\author[0000-0001-8638-0320]{Andrew W. Howard}
\affiliation{Department of Astronomy, California Institute of
  Technology, Pasadena, CA 91125, USA}


\begin{abstract}

Given the vast number of stars that exist within binary systems, it
remains important to explore the effect of binary star environments on
the formation and evolution of exoplanetary systems. Nearby binaries
provide opportunities to characterize their properties and orbits
through a combination of radial velocities, astrometry, and direct
imaging. Eta Cassiopeiae is a bright, well-known binary system for
which recent observations have provided greatly improved stellar
masses and orbital parameters. We present additional radial velocity
data that are used to perform an injection-recovery analysis for
potential planetary signatures. We further provide a detailed
dynamical study that explores the viability of planetary orbits
throughout the system. Our combined analysis shows that giant planets
are significantly ruled out for the system, and indeed no planetary
orbits are viable beyond $\sim$8~AU of the primary star. However,
terrestrial planets may yet exist within the Habitable Zone where
orbits can remain long-term stable. We discuss the implications of
these results, highlighting the effect of wide binary companions on
giant planet formation, and the consequences for occurrence rates and
planetary habitability.

\end{abstract}



\section{Introduction}
\label{intro}

A crucial component of studying exoplanetary systems is the
characterization of the stellar components within the system. The
stars determine the radiation environment of the planets, the
photochemistry of the planetary atmospheres, and the gravitational
well on which the dynamics of the system are overlaid. Numerous
surveys have been conducted whose aim is to characterize exoplanet
hosts and measure their fundamental properties
\citep{vonbraun2014,silvaaguirre2015,brewer2016b,petigura2017a,wittenmyer2020c},
the outcome of which are crucial in determining the properties of
their associated planets \citep{buchhave2014,fulton2018b}. However, a
substantial fraction of stellar systems consist of multiple
gravitationally bound stars
\citep{duquennoy1991a,duquennoy1991b,halbwachs2005,duchene2013b}.  The
presence of a stellar companion can have a profound effect on planet
formation and evolution \citep{ziegler2020,hirsch2021}, and also on
the planetary properties that are extracted \citep{ciardi2015a}.
Stellar companions have become an increasingly studied component of
known exoplanet systems, particularly as imaging techniques allow for
their detection and characterization
\citep{horch2014,wittrock2016,wittrock2017,matson2018,dalba2021b}. Additionally,
the increasing use of radial velocity (RV) data in conjunction with
precision astrometric techniques enable significantly improved orbits
of stellar binaries to be determined \citep{pearce2020}.

Eta Cassiopeiae (hereafter Eta~Cas) is a well-known binary star system
consisting of early G and late K main sequence stars in a $\sim$470~yr
orbit. The Eta Cas system is relatively close (5.92~pc) and the orbit
of the binary has been studied for over a century
\citep{doberck1876,vanderkamp1948,strand1969}. Over recent decades, it
has been the target of RV exoplanet surveys \citep{rosenthal2021}, and
more recently it has been observed with the Gaia mission, acquiring
valuable parallax and astrometric measurements \citep{brown2021}. The
NN-EXPLORE Exoplanet Investigations with Doppler Spectroscopy (NEID)
instrument \citep{schwab2016,robertson2019} observed Eta Cas via the
NEID Earth Twin Survey \citep{gupta2021}, providing RVs that were
merged with those from several other instruments
\citep{giovinazzi2025}. These RV data were combined with Gaia
astrometric data to determine a highly improved orbit and stellar
masses for Eta Cas compared with previous estimates, greatly advancing
our knowledge of the system. Such stellar binary characterizations
enable an assessment of detection limits for planets that may be
present, and also dynamical constraints due to the presence of the
binary companion, such as has been carried out for the Alpha Centauri
system \citep{quarles2016}. The orbit and nature of the binary
companion can also have important consequences for potentially
habitable environments within the system and the extent of the
Habitable Zone (HZ) around each of the stellar components
\citep{kasting1993a,kane2012a,kopparapu2013a,kopparapu2014,kane2016c,hill2018,hill2023}.

In this paper, we present an analysis of the Eta Cas system that
explores limitations imposed by both the RV data and dynamical effects
on the presence of possible planets within the system. The combination
of these approaches are used to demonstrate that the Eta Cas system is
unlikely to host any giant planets, although terrestrial planet orbits
within the HZ remain viable. Section~\ref{arch} describes the
architecture of the binary system and the extent of the
HZ. Section~\ref{rv} presents the revised RV data for the system,
including new Keck/HIRES data, and provides the results from an
injection-recovery test for planetary signatures. Section~\ref{dyn}
details the methodology and results for our dynamical simulation of
potentially stable planetary orbits within the system. The overall
results and their implications are discussed in
Section~\ref{discussion}, along with consequences for planet formation
and occurrence rates in binary systems. Section~\ref{conclusions}
provides concluding remarks and suggestions for future work.


\section{System Architecture}
\label{arch}


\subsection{The Stellar Binary}
\label{binary}

Eta~Cas is a binary star system at a distance of 5.92~pc and has
numerous designations, including HD~4614, HIP~3821, and HR~219. Based
on the astrometric and RV analysis provided by \citet{giovinazzi2025},
the mass of the primary star (A) is $M_{\star,A} = 1.0258$~$M_\odot$
and the mass of the secondary star (B) is $M_{\star,B} =
0.5487$~$M_\odot$. Their analysis also greatly improved the orbit of
the binary with the following parameters: orbital period $P =
472.2$~yrs, semi-major axis $a = 70.55$~AU, eccentricity $e =
0.49416$, and argument of periastron $\omega = 88.34\degr$. These
orbital parameters result in periastron and apastron separations of
35.79~AU and 105.41~AU, respectively. An isochrone analysis performed
by \citet{boyajian2012a} provided an estimated age for the system of
$5.4\pm0.9$~Gyr, placing it as $\sim$20\% older than the Sun.

Regarding the further required stellar properties of Eta~Cas~A, we
adopt the effective temperature provided by the GAIA DR3 data release
of $T_\mathrm{eff} = 5726$~K \citep{brown2021}, and the stellar radius
of $R_\star = 1.0336$~$R_\odot$ calculated by
\citet{giovinazzi2025}. From these stellar parameters, we calculate a
luminosity for Eta~Cas~A of $L_\star = 1.03$~$L_\odot$, where we note
that this is less than the value of 1.2321~$L_\odot$ estimated by
\citet{boyajian2012a}, primarily due to the lower GAIA DR3 effective
temperature. The existence of these parameter discrepancies for such a
bright (primary star $V = 3.44$), nearby star emphasizes the need for
continued stellar characterization, particularly for those in binary
systems and/or potential exoplanet hosts.


\subsection{The Habitable Zone}
\label{hz}

The HZ within planetary systems is a useful tool for the purposes of
target selection for follow-up observations that seek to further
characterize potentially habitable environments within the system
\citep{kasting1993a,kane2012a,kopparapu2013a,kopparapu2014,kane2016c,hill2018,hill2023}. Given
the relatively large fraction of binary stars
\citep{duquennoy1991a,duquennoy1991b,halbwachs2005,duchene2013b}, the
presence of an additional star can influence the HZ boundaries. These
modified HZ boundaries have been studied in the case of circumbinary
planets, or P-type orbits
\citep{haghighipour2013c,kane2013a,cukier2019}, and also for wide
binaries where planetary orbits around a single stellar component are
possible, or S-type orbits
\citep{eggl2012,kaltenegger2013b,wang2017d}. The Eta Cas system falls
into the latter category, in which the secondary star passes as close
as 35.79~AU of the primary star every 472.2~yrs (see
Section~\ref{binary}). However, since the secondary is a mid K-type
star and the closest separation between the stars is larger than
Neptune's orbit is from the Sun, the presence of the secondary has a
negligible effect on the HZ of the primary.

To calculate the HZ boundaries of Eta~Cas~A, we adopt the methodology
described by \citet{kopparapu2013a,kopparapu2014}, which provides a
polynomial fit to 1-D climate modeling results for a range of main
sequence spectral types. The calculations results in two sets of HZ
boundaries: the conservative HZ (CHZ), based on runaway and maximum
greenhouse limits, and the optimistic HZ (OHZ), based on empirical
evidence regarding past surface water on Venus and Mars
\citep{kane2016c}. Using the stellar properties reported in
Section~\ref{binary}, we calculate the CHZ and OHZ regions to be
0.968--1.710~AU and 0.764--1.803~AU, respectively. The long-term
monitoring of the star, combined with the dynamical effects of the
binary stellar companion, will determine the likelihood of undetected
planets being present within the HZ.


\section{Radial Velocity Limits on Planetary Companions}
\label{rv}

Here we describe the acquisition of additional RV data, and our
analysis of the limits imposed for possible planets within the system.


\subsection{Radial Velocity Data}
\label{data}

Since Eta Cas is such a bright system, it has been observed with
multiple facilities over three decades. We include here data from the
Hamilton Echelle Spectrograph on the 3.0m Shane telescope at Lick
Observatory \citep{vogt1987b,fischer2014a}, the (post-upgraded) HIRES
echelle spectrograph on the Keck I telescope \citep{vogt1994}, the
Levy spectrometer on the Automated Planet Finder (APF)
\citep{radovan2014,vogt2014a}, and the NEID instrument on the 3.5m
WIYN telescope at Kitt Peak National Observatory
\citep{schwab2016,robertson2019}. The HIRES data include 5 years of
additional data acquired since the last data release by
\citet{rosenthal2021}. A subset of the RV data, including facility,
times of observation, relative RVs, and associated errors, are shown
in Table~\ref{tab:rv}. In total, 945 RV measurements are included in
our full dataset that span a period of $\sim$30~years.

\begin{deluxetable}{lccc}
  \tablewidth{0pc}
  \tablecaption{\label{tab:rv} HD~4614 radial velocities.}
  \tablehead{
    \colhead{Instrument} &
    \colhead{Date} &
    \colhead{RV} &
    \colhead{$\sigma$} \\
    \colhead{} &
    \colhead{(BJD -- 2440000)} &
    \colhead{(m/s)} &
    \colhead{(m/s)}
  }
  \startdata
Hamilton & 9969.89106 & -12.48 & 1.19 \\
Hamilton & 10013.76518 & -8.53 & 1.40 \\
Hamilton & 10299.93750 & -10.67 & 1.96 \\
Hamilton & 10326.83496 & -8.33 & 2.03 \\
Hamilton & 10421.72526 & 1.39 & 1.74 \\
Hamilton & 10437.64258 & -4.23 & 1.70 \\
Hamilton & 10441.62305 & 0.84 & 1.91 \\
Hamilton & 10640.96368 & 2.30 & 1.38 \\
Hamilton & 10645.99108 & -1.20 & 1.54 \\
Hamilton & 10655.96094 & 2.03 & 2.12 \\
Hamilton & 10657.01269 & -4.36 & 2.16 \\
Hamilton & 10681.95703 & -2.70 & 2.14 \\
Hamilton & 10767.65625 & -0.61 & 2.30 \\
Hamilton & 10793.67090 & -2.74 & 1.85 \\
Hamilton & 10794.71777 & 3.03 & 2.55 \\
Hamilton & 11131.74121 & 15.30 & 2.65 \\
Hamilton & 11175.64356 & 8.47 & 2.60 \\
Hamilton & 11417.00000 & -0.07 & 2.42 \\
Hamilton & 11446.00781 & 13.48 & 3.15 \\
Hamilton & 13217.94531 & 31.89 & 2.22 \\
Hamilton & 13392.62581 & 33.94 & 1.94 \\
HIRES & 14838.86585 & -24.25 & 1.43 \\
HIRES & 14866.70650 & -28.10 & 1.70 \\
HIRES & 14866.70783 & -25.70 & 1.52 \\
HIRES & 14984.12244 & -24.89 & 1.32 \\
HIRES & 14985.12109 & -30.06 & 1.43 \\
HIRES & 14985.12164 & -27.95 & 1.35 \\
HIRES & 14985.12218 & -28.30 & 1.49 \\
HIRES & 14986.12193 & -27.15 & 1.42 \\
HIRES & 14986.12244 & -26.72 & 1.36 \\
  \enddata
\tablenotetext{}{The full data set is available online.}
\end{deluxetable}

\begin{figure*}
  \begin{center}
    \includegraphics[width=16.0cm]{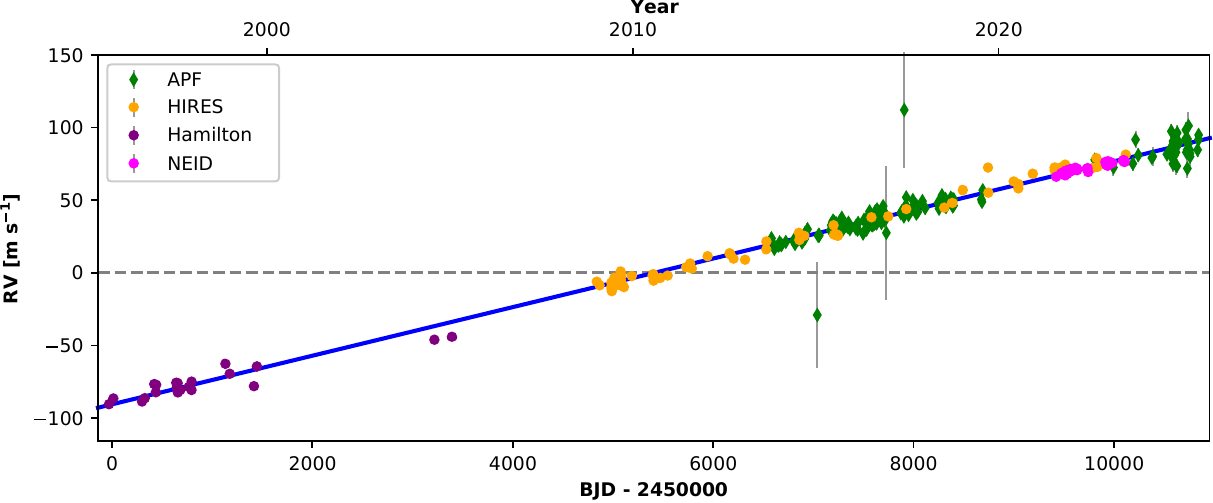}
  \end{center}
  \caption{The RV data for Eta Cas plotted against the observation
    time, including data from the APF, HIRES, Hamilton, and NEID
    instruments (see details in Section~\ref{data}). We include a
    linear fit to the observed RV trend caused by the binary orbit.}
  \label{fig:rvs}
\end{figure*}

\begin{figure}
  \includegraphics[width=8.5cm]{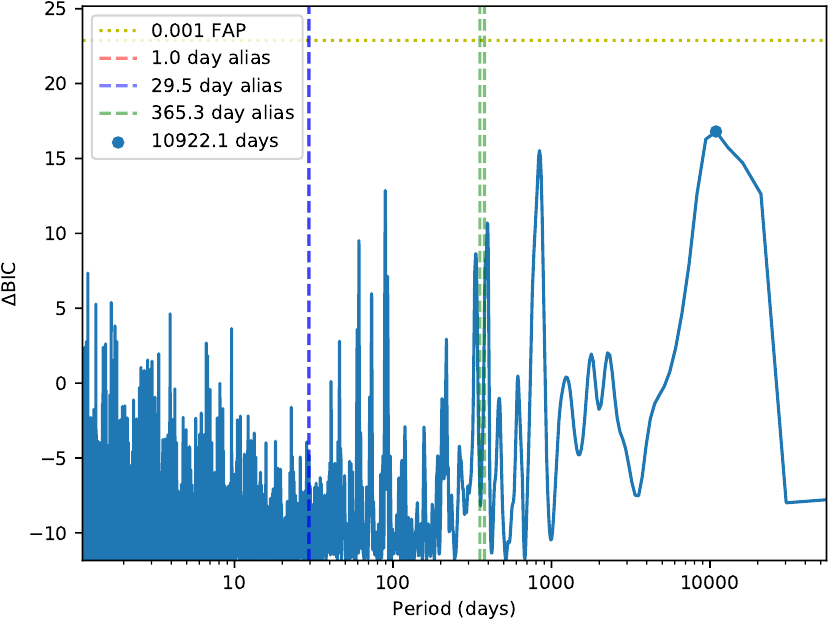}
  \caption{Power spectrum from a frequency analysis of the RV data
    shown in Figure~\ref{fig:rvs}. Significant observational alias
    locations (diurnal, lunar, annual) are shown as vertical dashed
    lines.}
  \label{fig:pgram}
\end{figure}

The RV data were analyzed using {\sc RVSearch}
\citep{fulton2018a,rosenthal2021}, an iterative planet-searching tool,
to search for periodic signals that may be associated with planets
within the system. We searched between a minimum period of 1 day and a
maximum period of 5 times the baseline of our combined dataset, which
is approximately 50,000 days, while allowing the inclusion of
systematic velocity offsets between different instruments and linear
or quadratic trend signatures in the search. We set a false alarm
probability (FAP) threshold of 0.1\%, below which periodic signals
detected by the search algorithm are considered to be statistically
significant. Shown in Figure~\ref{fig:rvs} are all of the RVs from
Table~\ref{tab:rv} along with a linear trend fit to the data. The
linear trend is the result of the binary stellar companion to Eta Cas,
and no significant periodic signatures were detected in the data due
to potential planets.

\citet{giovinazzi2025} note in their analysis the presence of a
868~day signal that may be attributed to a bound companion with a mass
as low as 22~$M_\oplus$. Shown in Figure~\ref{fig:pgram} is the power
spectrum resulting from our frequency analysis of the RV data, which
contains an additional 5 years of HIRES data compared with those used
by \citet{giovinazzi2025}. The horizontal dotted line indicates the
0.001 (0.1\%) false-alarm probability (FAP) threshold, and the
vertical dashed lines show the location of aliases resulting from the
diurnal, annual, and lunar cycles. Our periodogram shows the presence
of a periodic signature located at $\sim$841~days, equivalent to a
semi-major axis of 1.759~AU based on the stellar properties described
in Section~\ref{binary}. However, the power of the peak lies well
below the FAP threshold, and thus we conclude that there are no
significant periodic signals present in the data.


\subsection{Injection-Recovery Analysis}
\label{inj}

\begin{figure}
  \includegraphics[width=8.5cm]{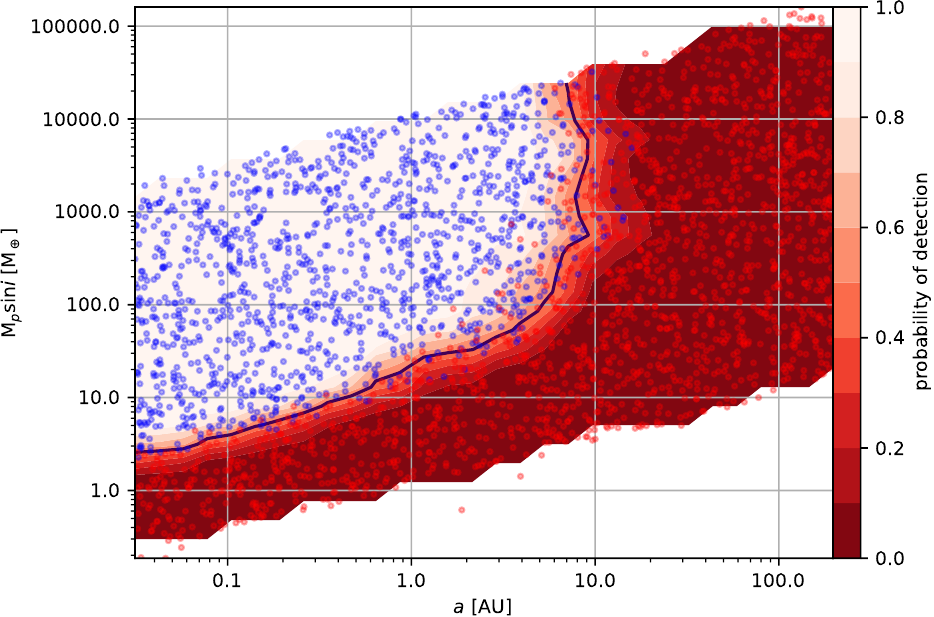}
  \caption{Results of the injection-recovery test to assess the
    sensitivity of our RV data to planetary signatures as a function
    of planetary mass ($M_p \sin i$) and semi-major axis ($a$). The
    blue dots indicate successfully recovered injected planetary
    signatures and the red dots indicate unrecovered planets. The
    color scale corresponding to the probability contours of detecting
    a planet of a given mass and semi-major axis is shown on the right
    vertical axis.}
  \label{fig:inj}
\end{figure}

Using our RV data (see Section~\ref{data}) and the {\sc RVSearch}
tool, we conducted an injection-recovery test to explore the
sensitivity of our dataset to planetary signatures. This was done by
injecting 3000 synthetic planets of different minimum masses and
orbital periods that were drawn from log-uniform distribution and
orbital eccentricity from the beta distribution
\citep{kipping2013b}. The details regarding the methodology for this
process, including the criteria for planetary signature recovery, are
described by \citet{howard2016}. The recovery rate from this
injection-recovery process forms a completeness contour that informs
the sensitivity of our RV data to the minimum mass versus semi-major
axis space, shown in Figure~\ref{fig:inj}. The blue and red dots
indicate recovered and missed injected signals, respectively, and our
data are not sensitive to objects that lie below the black solid line,
which represents the 50\% detection probability threshold. These
results show that the precision of our combined dataset does not allow
the detection of terrestrial planets in any part of the semi-major
axis space. However, we can rule out the presence of gas giant planets
within $\sim$8~AU of the primary star as our data would have detected
such a planetary presence if they were there. Therefore, low-mass
terrestrial planets orbiting inside the HZ and cold gas giant planets
in the outer regime of this system cannot be ruled out by the RV data.

\section{Dynamical Limits on Planetary Companions}
\label{dyn}

Here we present our dynamical analysis of the system and consequences
for planetary orbital stability.


\subsection{Dynamical Simulation Methodology}
\label{methods}

The primary purpose of our dynamical analysis is to quantify the
gravitational perturbations of the secondary star on potential
planetary orbits around the primary star. \citet{holman1999} provide
empirically derived formula for calculating the critical semi-major
axis for a circumstellar orbit, beyond which the orbit is likely to be
unstable. Using the binary star properties provided by
\citet{giovinazzi2025} and described in Section~\ref{binary}, we
calculate a critical semi-major axis for Eta~Cas~A of 9.91~AU, which
is beyond the limits imposed by the RV data provided in
Section~\ref{rv}.

To assess the viability of planetary orbits around the primary star,
we carried out a series of dynamical simulations using the REBOUND
N-body integrator package \citep{rein2012a}, which applies the
symplectic integrator WHFast \citep{rein2015c}. Each simulation
tracked the orbital evolution of an Earth-mass planet injected into a
circular orbit that is coplanar with the orbit of the binary. We
explored the semi-major axis range of 0.5--11.0~AU, in steps of
0.0015~AU, for a total of 7001 simulations. This semi-major axis range
is equivalent to an orbital period range of 127.5--13155~days. This
semi-major axis range was chosen to encompass the full HZ region,
described in Section~\ref{hz}, and also extend beyond the critical
semi-major axis limit calculated from the \citet{holman1999} formula.
\citet{rabl1988} concluded that an integration time equivalent to 300
binary star periods is sufficient to determine stability
boundaries. Thus, each of our semi-major axis starting locations were
integrated for $10^7$~years, recording orbital elements every
100~years, and with a time step of 5~days to ensure adequate time
resolution of planetary perturbations due to the secondary star. At
the conclusion of each simulation, we calculated the maximum
eccentricity of the planet and the percentage of the integration for
which the planet remained within the system. Loss of the planet
generally means that it has either been ejected from the system or
consumed by the gravitational well of the primary star.


\subsection{Dynamical Stability Results}
\label{stab}

\begin{figure*}
  \begin{center}
    \includegraphics[width=16.05cm]{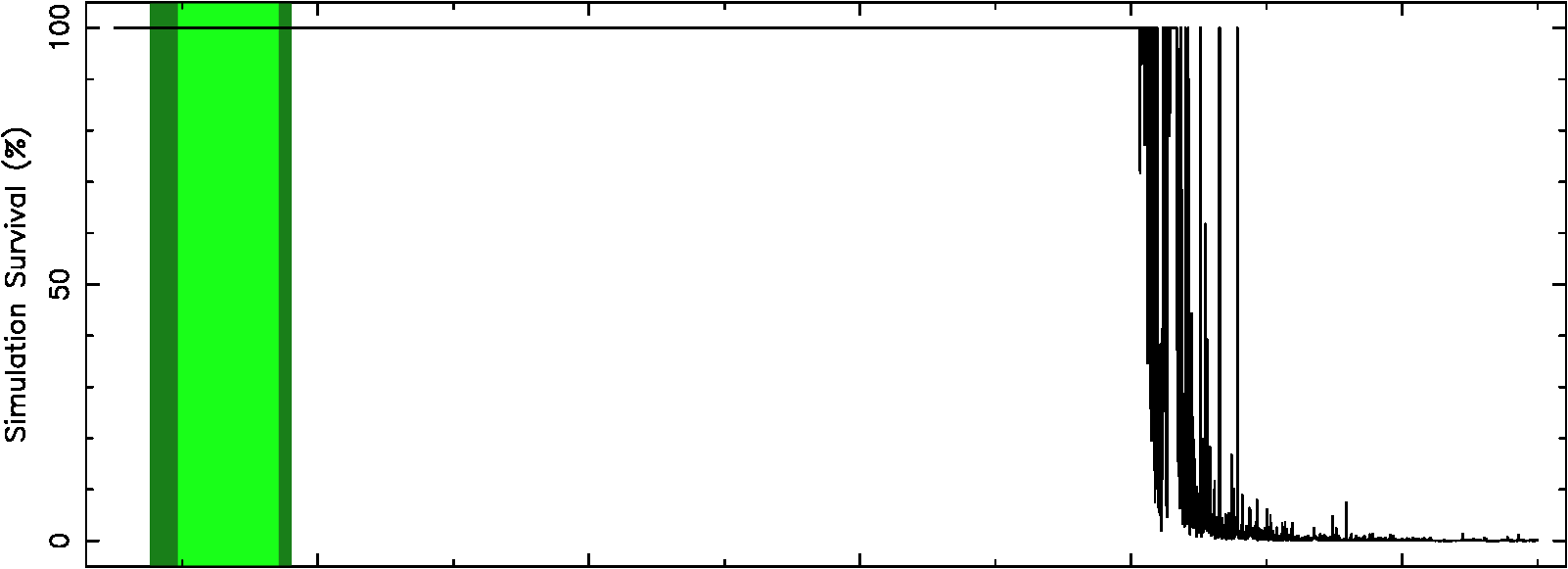} \\
    \includegraphics[width=16.0cm]{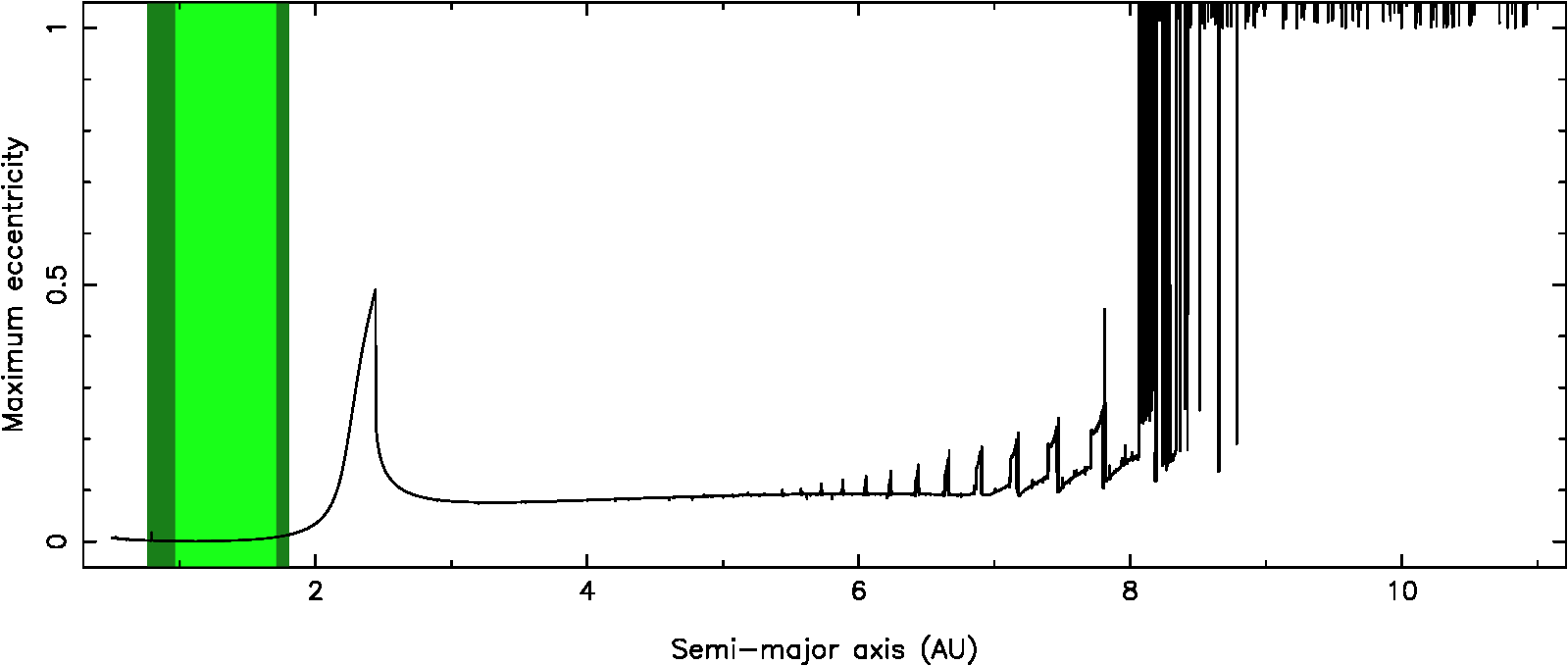}
  \end{center}
  \caption{Results of the dynamical simulation for the Eta~Cas system
    as a function of semi-major axis of the injected planet. The top
    panel provides the percentage survival calculations for the
    planet, and the bottom panel shows the maximum eccentricity of the
    planet. The extent of the HZ is shown in green, where the light
    green and dark green regions indicate the CHZ and OHZ,
    respectively.}
  \label{fig:sim}
\end{figure*}

The results of our dynamical simulations are shown in
Figure~\ref{fig:sim}. For each semi-major axis, the percentage of the
full integration for which the planet survived is shown in the top
panel, and the maximum recorded eccentricity of the planet is shown in
the bottom panel. The extent of the HZ region, as calculated in
Section~\ref{hz}, is also shown, where the light green and dark green
regions indicate the CHZ and OHZ, respectively. The survival results
in the top panel reveal that the closest approach of the binary
companion ($\sim$35~AU; see Section~\ref{binary}) causes orbital
instability for the injected planet for semi-major axis values beyond
$\sim$8~AU. This dynamical limit is less than that suggested by
\citet{holman1999}, likely the result of the relatively high
eccentricity of the binary. Although there are a plethora of possible
stable planetary orbits between 2--8~AU, these locations generally
have induced eccentricities of $\sim$0.1, and can be as high as
$\sim$0.5, produced by the interaction of orbital precession with
resonance locations of the binary \citep{touma2015}. Furthermore, the
dynamical limits presented here match well with the RV limits
described in Section~\ref{inj} that severely constrain the presence of
giant planets out to 8~AU. However, the good news is that the HZ
appears to be a dynamically stable region with little to no
significant orbital perturbations that excite the orbital
eccentricity. Thus, there remains the possibility of terrestrial
planets being present in orbit around the primary star, and especially
within the HZ region.


\section{Discussion}
\label{discussion}

The exploration of planetary systems in the context of stellar
binaries remains a fascinating topic, particularly in the context of
single-star systems, such as the solar system. Exoplanet surveys have
found a higher occurrence rate of planets in single-star than
multiple-star systems with separations smaller than $\sim$1500~AU
\citep{wang2014b}. Further surveys have identified a difference in the
inner versus outer system planet occurrence rates, depending on the
stellar multiplicity \citep{hirsch2021}. Such differences in planetary
architectures point to the fundamental planet formation and evolution
effects of a wide binary companion \citep{takeda2008,petrovich2015a},
and the consequences of the evolving nature of the binary orbit itself
due to galactic perturbations \citep{kaib2013}. Even the solar system
is not immune to such effects, as various simulations of the solar
neighborhood indicate the critical role that past and future stellar
encounters will have in shaping the architecture of the outer solar
system orbits
\citep{zink2020c,brown2022a,raymond2024a,kaib2025}. Indeed, stellar
flybys can have an extremely disruptive effect on planet formation and
evolution in single-star systems, including planets within the HZ
\citep{malmberg2007b,malmberg2011,hao2013,kane2018b}, and likely
contributes to the free-floating planet population
\citep{veras2012b}. Although we do not incorporate the external
influences that may have acted upon the orbit of the Eta~Cas binary
system over $10^7$~year time scales, our results show that the
dynamics of the binary are sufficient to eliminate planetary orbits
outside of 8~AU of the primary. Inside of 8~AU, the situation is more
complicated. The limits imposed by our RV data analysis suggest that
there are no giant planets in this inner region, consistent with
suggestions that binary stars may truncate disk life times and thus
the opportunity to form giant planets \citep{kraus2012b}. The
dynamical results show that the presence of the binary is also felt
inside of 8~AU, with perturbations that can excite orbital
eccentricities as close as 2~AU.

The lack of giant planets within the Eta~Cas system is similar to the
planetary constraints found for Beta~CVn \citep{kane2020b}, which is a
single-star and a target of astrobiological significance, due to both
its similarity and proximity to the Sun
\citep{portodemello2006,turnbull2015}. Assessing the orbital stability
within the HZ of systems, both with and without giant planets and
stellar companions, is a crucial step in considering targets for
potential follow-up observations that aim to detect and/or
characterize HZ terrestrial planets
\citep{kopparapu2010,kane2020b,kane2022a,kane2023c}. Such work is of
particular relevance at the present time, as significant efforts are
being directed toward the selection of nearby stellar targets for a
possible future flagship direct imaging mission
\citep{laliotis2023,harada2024b,tuchow2024}. The presence of other
planets in the system, especially those giant planets whose eccentric
orbits pass near or through the HZ, can effectively eliminate those
systems as viable search targets for HZ terrestrial planets
\citep{kane2024d,kane2024e}. Similarly, a binary companion can
eradicate stable HZ orbits, or otherwise induce considerable
disruptive climate effects on HZ terrestrial planets, including
runaway glaciation \citep{quarles2022}. In the case of Eta~Cas, the
periastron separation of the secondary component is sufficiently large
to leave the HZ relatively untouched, making Eta~Cas an interesting
target for further RV and direct imaging studies.


\section{Conclusions}
\label{conclusions}

The Eta~Cas system is one of the most well-known and historically
documented binaries, whose observations in the era of precision
astrometry and RVs has produced an exquisitely defined orbit. The
revised orbit, along with additional RV data that extends our
observational baseline, has allowed us to conduct a detailed analysis
on how the presence of the binary stellar companion has influenced the
possibility of planets within the system. The injection of planetary
signatures into the RV data show that, were there giant planets
orbiting the stellar primary, our RV data should be sufficient to
detect those signatures out to $\sim$8~AU. From the other direction,
the results of our exhaustive suite of dynamical simulations have
ruled out any stable orbits beyond $\sim$8~AU, effectively eliminating
the likelihood of giant planets orbiting the primary star. Although
the dynamical effects of the binary star can excite the eccentricities
of any planetary orbits as close as $\sim$2~AU to the primary, the HZ
region remains quite stable, experiencing little to no ill effects
from the binary companion. Thus, the HZ of the primary remains a
viable home for terrestrial planets that may be the subject of further
RV and future direct imaging campaigns. Our results demonstrate the
power of combining long baseline RV data with dynamical analyses that
provide a more complete picture of planetary detection limits.


\section*{Acknowledgements}

The authors thank the anonymous referee for their feedback that
improved the manuscript. We gratefully acknowledge the efforts and
dedication of the Keck Observatory staff for support of HIRES and
remote observing. We recognize and acknowledge the cultural role and
reverence that the summit of Maunakea has within the indigenous
Hawaiian community. We are deeply grateful to have the opportunity to
conduct observations from this mountain. We thank Ken and Gloria Levy,
who supported the construction of the Levy Spectrometer on the
Automated Planet Finder. We thank the University of California and
Google for supporting Lick Observatory and the UCO staff for their
dedicated work scheduling and operating the telescopes of Lick
Observatory. This research has made use of the Habitable Zone Gallery
at hzgallery.org. The results reported herein benefited from
collaborations and/or information exchange within NASA's Nexus for
Exoplanet System Science (NExSS) research coordination network
sponsored by NASA's Science Mission Directorate.


\software{REBOUND \citep{rein2012a}, RadVel \citep{fulton2018a},
  RVSearch \citep{rosenthal2021}}



\begin{thebibliography}{}
\expandafter\ifx\csname natexlab\endcsname\relax\def\natexlab#1{#1}\fi
\providecommand{\url}[1]{\href{#1}{#1}}
\providecommand{\dodoi}[1]{doi:~\href{http://doi.org/#1}{\nolinkurl{#1}}}
\providecommand{\doeprint}[1]{\href{http://ascl.net/#1}{\nolinkurl{http://ascl.net/#1}}}
\providecommand{\doarXiv}[1]{\href{https://arxiv.org/abs/#1}{\nolinkurl{https://arxiv.org/abs/#1}}}

\bibitem[{{Boyajian} {et~al.}(2012){Boyajian}, {McAlister}, {van Belle},
  {Gies}, {ten Brummelaar}, {von Braun}, {Farrington}, {Goldfinger}, {O'Brien},
  {Parks}, {Richardson}, {Ridgway}, {Schaefer}, {Sturmann}, {Sturmann},
  {Touhami}, {Turner}, \& {White}}]{boyajian2012a}
{Boyajian}, T.~S., {McAlister}, H.~A., {van Belle}, G., {et~al.} 2012, \apj,
  746, 101, \dodoi{10.1088/0004-637X/746/1/101}

\bibitem[{{Brewer} {et~al.}(2016){Brewer}, {Fischer}, {Valenti}, \&
  {Piskunov}}]{brewer2016b}
{Brewer}, J.~M., {Fischer}, D.~A., {Valenti}, J.~A., \& {Piskunov}, N. 2016,
  \apjs, 225, 32, \dodoi{10.3847/0067-0049/225/2/32}

\bibitem[{{Brown} \& {Rein}(2022)}]{brown2022a}
{Brown}, G., \& {Rein}, H. 2022, \mnras, 515, 5942,
  \dodoi{10.1093/mnras/stac1763}

\bibitem[{{Buchhave} {et~al.}(2014){Buchhave}, {Bizzarro}, {Latham},
  {Sasselov}, {Cochran}, {Endl}, {Isaacson}, {Juncher}, \&
  {Marcy}}]{buchhave2014}
{Buchhave}, L.~A., {Bizzarro}, M., {Latham}, D.~W., {et~al.} 2014, \nat, 509,
  593, \dodoi{10.1038/nature13254}

\bibitem[{{Ciardi} {et~al.}(2015){Ciardi}, {Beichman}, {Horch}, \&
  {Howell}}]{ciardi2015a}
{Ciardi}, D.~R., {Beichman}, C.~A., {Horch}, E.~P., \& {Howell}, S.~B. 2015,
  \apj, 805, 16, \dodoi{10.1088/0004-637X/805/1/16}

\bibitem[{{Cukier} {et~al.}(2019){Cukier}, {Kopparapu}, {Kane}, {Welsh},
  {Wolf}, {Kostov}, \& {Haqq-Misra}}]{cukier2019}
{Cukier}, W., {Kopparapu}, R.~k., {Kane}, S.~R., {et~al.} 2019, \pasp, 131,
  124402, \dodoi{10.1088/1538-3873/ab50cb}

\bibitem[{{Dalba} {et~al.}(2021){Dalba}, {Kane}, {Howell}, {Horch}, {Li},
  {Hirsch}, {Burt}, {Brandt}, {Mo{\v{c}}nik}, {Henry}, {Everett}, {Rosenthal},
  \& {Howard}}]{dalba2021b}
{Dalba}, P.~A., {Kane}, S.~R., {Howell}, S.~B., {et~al.} 2021, \aj, 161, 123,
  \dodoi{10.3847/1538-3881/abd6ed}

\bibitem[{{Doberck}(1876)}]{doberck1876}
{Doberck}, W. 1876, Astronomische Nachrichten, 88, 45,
  \dodoi{10.1002/asna.18760880303}

\bibitem[{{Duch{\^e}ne} \& {Kraus}(2013)}]{duchene2013b}
{Duch{\^e}ne}, G., \& {Kraus}, A. 2013, \araa, 51, 269,
  \dodoi{10.1146/annurev-astro-081710-102602}

\bibitem[{{Duquennoy} \& {Mayor}(1991)}]{duquennoy1991b}
{Duquennoy}, A., \& {Mayor}, M. 1991, \aap, 248, 485

\bibitem[{{Duquennoy} {et~al.}(1991){Duquennoy}, {Mayor}, \&
  {Halbwachs}}]{duquennoy1991a}
{Duquennoy}, A., {Mayor}, M., \& {Halbwachs}, J.-L. 1991, \aaps, 88, 281

\bibitem[{{Eggl} {et~al.}(2012){Eggl}, {Pilat-Lohinger}, {Georgakarakos},
  {Gyergyovits}, \& {Funk}}]{eggl2012}
{Eggl}, S., {Pilat-Lohinger}, E., {Georgakarakos}, N., {Gyergyovits}, M., \&
  {Funk}, B. 2012, \apj, 752, 74, \dodoi{10.1088/0004-637X/752/1/74}

\bibitem[{{Fischer} {et~al.}(2014){Fischer}, {Marcy}, \&
  {Spronck}}]{fischer2014a}
{Fischer}, D.~A., {Marcy}, G.~W., \& {Spronck}, J. F.~P. 2014, \apjs, 210, 5,
  \dodoi{10.1088/0067-0049/210/1/5}

\bibitem[{{Fulton} \& {Petigura}(2018)}]{fulton2018b}
{Fulton}, B.~J., \& {Petigura}, E.~A. 2018, \aj, 156, 264,
  \dodoi{10.3847/1538-3881/aae828}

\bibitem[{{Fulton} {et~al.}(2018){Fulton}, {Petigura}, {Blunt}, \&
  {Sinukoff}}]{fulton2018a}
{Fulton}, B.~J., {Petigura}, E.~A., {Blunt}, S., \& {Sinukoff}, E. 2018, \pasp,
  130, 044504, \dodoi{10.1088/1538-3873/aaaaa8}

\bibitem[{{Gaia Collaboration} {et~al.}(2021){Gaia Collaboration}, {Brown},
  {Vallenari}, {Prusti}, {de Bruijne}, {Babusiaux}, {Biermann}, {Creevey},
  {Evans}, {Eyer}, {Hutton}, {Jansen}, {Jordi}, {Klioner}, {Lammers},
  {Lindegren}, {Luri}, {Mignard}, {Panem}, {Pourbaix}, {Randich}, {Sartoretti},
  {Soubiran}, {Walton}, {Arenou}, {Bailer-Jones}, {Bastian}, {Cropper},
  {Drimmel}, {Katz}, {Lattanzi}, {van Leeuwen}, {Bakker}, {Cacciari},
  {Casta{\~n}eda}, {De Angeli}, {Ducourant}, {Fabricius}, {Fouesneau},
  {Fr{\'e}mat}, {Guerra}, {Guerrier}, {Guiraud}, {Jean-Antoine Piccolo},
  {Masana}, {Messineo}, {Mowlavi}, {Nicolas}, {Nienartowicz}, {Pailler},
  {Panuzzo}, {Riclet}, {Roux}, {Seabroke}, {Sordo}, {Tanga}, {Th{\'e}venin},
  {Gracia-Abril}, {Portell}, {Teyssier}, {Altmann}, {Andrae}, {Bellas-Velidis},
  {Benson}, {Berthier}, {Blomme}, {Brugaletta}, {Burgess}, {Busso}, {Carry},
  {Cellino}, {Cheek}, {Clementini}, {Damerdji}, {Davidson}, {Delchambre},
  {Dell'Oro}, {Fern{\'a}ndez-Hern{\'a}ndez}, {Galluccio}, {Garc{\'\i}a-Lario},
  {Garcia-Reinaldos}, {Gonz{\'a}lez-N{\'u}{\~n}ez}, {Gosset}, {Haigron},
  {Halbwachs}, {Hambly}, {Harrison}, {Hatzidimitriou}, {Heiter},
  {Hern{\'a}ndez}, {Hestroffer}, {Hodgkin}, {Holl}, {Jan{\ss}en}, {Jevardat de
  Fombelle}, {Jordan}, {Krone-Martins}, {Lanzafame}, {L{\"o}ffler}, {Lorca},
  {Manteiga}, {Marchal}, {Marrese}, {Moitinho}, {Mora}, {Muinonen}, {Osborne},
  {Pancino}, {Pauwels}, {Petit}, {Recio-Blanco}, {Richards}, {Riello},
  {Rimoldini}, {Robin}, {Roegiers}, {Rybizki}, {Sarro}, {Siopis}, {Smith},
  {Sozzetti}, {Ulla}, {Utrilla}, {van Leeuwen}, {van Reeven}, {Abbas}, {Abreu
  Aramburu}, {Accart}, {Aerts}, {Aguado}, {Ajaj}, {Altavilla}, {{\'A}lvarez},
  {{\'A}lvarez Cid-Fuentes}, {Alves}, {Anderson}, {Anglada Varela}, {Antoja},
  {Audard}, {Baines}, {Baker}, {Balaguer-N{\'u}{\~n}ez}, {Balbinot}, {Balog},
  {Barache}, {Barbato}, {Barros}, {Barstow}, {Bartolom{\'e}}, {Bassilana},
  {Bauchet}, {Baudesson-Stella}, {Becciani}, {Bellazzini}, {Bernet}, {Bertone},
  {Bianchi}, {Blanco-Cuaresma}, {Boch}, {Bombrun}, {Bossini}, {Bouquillon},
  {Bragaglia}, {Bramante}, {Breedt}, {Bressan}, {Brouillet}, {Bucciarelli},
  {Burlacu}, {Busonero}, {Butkevich}, {Buzzi}, {Caffau}, {Cancelliere},
  {C{\'a}novas}, {Cantat-Gaudin}, {Carballo}, {Carlucci}, {Carnerero},
  {Carrasco}, {Casamiquela}, {Castellani}, {Castro-Ginard}, {Castro Sampol},
  {Chaoul}, {Charlot}, {Chemin}, {Chiavassa}, {Cioni}, {Comoretto}, {Cooper},
  {Cornez}, {Cowell}, {Crifo}, {Crosta}, {Crowley}, {Dafonte}, {Dapergolas},
  {David}, {David}, {de Laverny}, {De Luise}, {De March}, {De Ridder}, {de
  Souza}, {de Teodoro}, {de Torres}, {del Peloso}, {del Pozo}, {Delbo},
  {Delgado}, {Delgado}, {Delisle}, {Di Matteo}, {Diakite}, {Diener},
  {Distefano}, {Dolding}, {Eappachen}, {Edvardsson}, {Enke}, {Esquej}, {Fabre},
  {Fabrizio}, {Faigler}, {Fedorets}, {Fernique}, {Fienga}, {Figueras},
  {Fouron}, {Fragkoudi}, {Fraile}, {Franke}, {Gai}, {Garabato},
  {Garcia-Gutierrez}, {Garc{\'\i}a-Torres}, {Garofalo}, {Gavras}, {Gerlach},
  {Geyer}, {Giacobbe}, {Gilmore}, {Girona}, {Giuffrida}, {Gomel}, {Gomez},
  {Gonzalez-Santamaria}, {Gonz{\'a}lez-Vidal}, {Granvik},
  {Guti{\'e}rrez-S{\'a}nchez}, {Guy}, {Hauser}, {Haywood}, {Helmi}, {Hidalgo},
  {Hilger}, {H{\l}adczuk}, {Hobbs}, {Holland}, {Huckle}, {Jasniewicz},
  {Jonker}, {Juaristi Campillo}, {Julbe}, {Karbevska}, {Kervella}, {Khanna},
  {Kochoska}, {Kontizas}, {Kordopatis}, {Korn}, {Kostrzewa-Rutkowska},
  {Kruszy{\'n}ska}, {Lambert}, {Lanza}, {Lasne}, {Le Campion}, {Le Fustec},
  {Lebreton}, {Lebzelter}, {Leccia}, {Leclerc}, {Lecoeur-Taibi}, {Liao},
  {Licata}, {Lindstr{\o}m}, {Lister}, {Livanou}, {Lobel}, {Madrero Pardo},
  {Managau}, {Mann}, {Marchant}, {Marconi}, {Marcos Santos}, {Marinoni},
  {Marocco}, {Marshall}, {Martin Polo}, {Mart{\'\i}n-Fleitas}, {Masip},
  {Massari}, {Mastrobuono-Battisti}, {Mazeh}, {McMillan}, {Messina},
  {Michalik}, {Millar}, {Mints}, {Molina}, {Molinaro}, {Moln{\'a}r},
  {Montegriffo}, {Mor}, {Morbidelli}, {Morel}, {Morris}, {Mulone}, {Munoz},
  {Muraveva}, {Murphy}, {Musella}, {Noval}, {Ord{\'e}novic}, {Orr{\`u}},
  {Osinde}, {Pagani}, {Pagano}, {Palaversa}, {Palicio}, {Panahi}, {Pawlak},
  {Pe{\~n}alosa Esteller}, {Penttil{\"a}}, {Piersimoni}, {Pineau}, {Plachy},
  {Plum}, {Poggio}, {Poretti}, {Poujoulet}, {Pr{\v{s}}a}, {Pulone}, {Racero},
  {Ragaini}, {Rainer}, {Raiteri}, {Rambaux}, {Ramos}, {Ramos-Lerate}, {Re
  Fiorentin}, {Regibo}, {Reyl{\'e}}, {Ripepi}, {Riva}, {Rixon}, {Robichon},
  {Robin}, {Roelens}, {Rohrbasser}, {Romero-G{\'o}mez}, {Rowell}, {Royer},
  {Rybicki}, {Sadowski}, {Sagrist{\`a} Sell{\'e}s}, {Sahlmann}, {Salgado},
  {Salguero}, {Samaras}, {Sanchez Gimenez}, {Sanna}, {Santove{\~n}a},
  {Sarasso}, {Schultheis}, {Sciacca}, {Segol}, {Segovia}, {S{\'e}gransan},
  {Semeux}, {Shahaf}, {Siddiqui}, {Siebert}, {Siltala}, {Slezak}, {Smart},
  {Solano}, {Solitro}, {Souami}, {Souchay}, {Spagna}, {Spoto}, {Steele},
  {Steidelm{\"u}ller}, {Stephenson}, {S{\"u}veges}, {Szabados}, {Szegedi-Elek},
  {Taris}, {Tauran}, {Taylor}, {Teixeira}, {Thuillot}, {Tonello}, {Torra},
  {Torra}, {Turon}, {Unger}, {Vaillant}, {van Dillen}, {Vanel}, {Vecchiato},
  {Viala}, {Vicente}, {Voutsinas}, {Weiler}, {Wevers}, {Wyrzykowski}, {Yoldas},
  {Yvard}, {Zhao}, {Zorec}, {Zucker}, {Zurbach}, \& {Zwitter}}]{brown2021}
{Gaia Collaboration}, {Brown}, A.~G.~A., {Vallenari}, A., {et~al.} 2021, \aap,
  649, A1, \dodoi{10.1051/0004-6361/202039657}

\bibitem[{{Giovinazzi} {et~al.}(2025){Giovinazzi}, {Blake}, {Robertson}, {Lin},
  {Gupta}, {Mahadevan}, {Wright}, {Bardalez Gagliuffi}, {Dong}, {Fernandes},
  {Fitzmaurice}, {Halverson}, {Kanodia}, {Logsdon}, {Luhn}, {McElwain},
  {Monson}, {Ninan}, {Rajagopal}, {Roy}, {Schwab}, {Stef{\'a}nsson}, {Terrien},
  {Eastman}, {Horner}, {Plavchan}, {Wang}, {Wilson}, \&
  {Wittenmyer}}]{giovinazzi2025}
{Giovinazzi}, M.~R., {Blake}, C.~H., {Robertson}, P., {et~al.} 2025, \aj, 170,
  52, \dodoi{10.3847/1538-3881/add922}

\bibitem[{{Gupta} {et~al.}(2021){Gupta}, {Wright}, {Robertson}, {Halverson},
  {Luhn}, {Roy}, {Mahadevan}, {Ford}, {Bender}, {Blake}, {Hearty}, {Kanodia},
  {Logsdon}, {McElwain}, {Monson}, {Ninan}, {Schwab}, {Stef{\'a}nsson}, \&
  {Terrien}}]{gupta2021}
{Gupta}, A.~F., {Wright}, J.~T., {Robertson}, P., {et~al.} 2021, \aj, 161, 130,
  \dodoi{10.3847/1538-3881/abd79e}

\bibitem[{{Haghighipour} \& {Kaltenegger}(2013)}]{haghighipour2013c}
{Haghighipour}, N., \& {Kaltenegger}, L. 2013, \apj, 777, 166,
  \dodoi{10.1088/0004-637X/777/2/166}

\bibitem[{{Halbwachs} {et~al.}(2005){Halbwachs}, {Mayor}, \&
  {Udry}}]{halbwachs2005}
{Halbwachs}, J.~L., {Mayor}, M., \& {Udry}, S. 2005, \aap, 431, 1129,
  \dodoi{10.1051/0004-6361:20041219}

\bibitem[{{Hao} {et~al.}(2013){Hao}, {Kouwenhoven}, \& {Spurzem}}]{hao2013}
{Hao}, W., {Kouwenhoven}, M.~B.~N., \& {Spurzem}, R. 2013, \mnras, 433, 867,
  \dodoi{10.1093/mnras/stt771}

\bibitem[{{Harada} {et~al.}(2024){Harada}, {Dressing}, {Kane}, \&
  {Ardestani}}]{harada2024b}
{Harada}, C.~K., {Dressing}, C.~D., {Kane}, S.~R., \& {Ardestani}, B.~A. 2024,
  \apjs, 272, 30, \dodoi{10.3847/1538-4365/ad3e81}

\bibitem[{{Hill} {et~al.}(2023){Hill}, {Bott}, {Dalba}, {Fetherolf}, {Kane},
  {Kopparapu}, {Li}, \& {Ostberg}}]{hill2023}
{Hill}, M.~L., {Bott}, K., {Dalba}, P.~A., {et~al.} 2023, \aj, 165, 34,
  \dodoi{10.3847/1538-3881/aca1c0}

\bibitem[{{Hill} {et~al.}(2018){Hill}, {Kane}, {Seperuelo Duarte}, {Kopparapu},
  {Gelino}, \& {Wittenmyer}}]{hill2018}
{Hill}, M.~L., {Kane}, S.~R., {Seperuelo Duarte}, E., {et~al.} 2018, \apj, 860,
  67, \dodoi{10.3847/1538-4357/aac384}

\bibitem[{{Hirsch} {et~al.}(2021){Hirsch}, {Rosenthal}, {Fulton}, {Howard},
  {Ciardi}, {Marcy}, {Nielsen}, {Petigura}, {de Rosa}, {Isaacson}, {Weiss},
  {Sinukoff}, \& {Macintosh}}]{hirsch2021}
{Hirsch}, L.~A., {Rosenthal}, L., {Fulton}, B.~J., {et~al.} 2021, \aj, 161,
  134, \dodoi{10.3847/1538-3881/abd639}

\bibitem[{{Holman} \& {Wiegert}(1999)}]{holman1999}
{Holman}, M.~J., \& {Wiegert}, P.~A. 1999, \aj, 117, 621,
  \dodoi{10.1086/300695}

\bibitem[{{Horch} {et~al.}(2014){Horch}, {Howell}, {Everett}, \&
  {Ciardi}}]{horch2014}
{Horch}, E.~P., {Howell}, S.~B., {Everett}, M.~E., \& {Ciardi}, D.~R. 2014,
  \apj, 795, 60, \dodoi{10.1088/0004-637X/795/1/60}

\bibitem[{{Howard} \& {Fulton}(2016)}]{howard2016}
{Howard}, A.~W., \& {Fulton}, B.~J. 2016, \pasp, 128, 114401,
  \dodoi{10.1088/1538-3873/128/969/114401}

\bibitem[{{Kaib} \& {Raymond}(2025)}]{kaib2025}
{Kaib}, N.~A., \& {Raymond}, S.~N. 2025, \icarus, 439, 116632,
  \dodoi{10.1016/j.icarus.2025.116632}

\bibitem[{{Kaib} {et~al.}(2013){Kaib}, {Raymond}, \& {Duncan}}]{kaib2013}
{Kaib}, N.~A., {Raymond}, S.~N., \& {Duncan}, M. 2013, \nat, 493, 381,
  \dodoi{10.1038/nature11780}

\bibitem[{{Kaltenegger} \& {Haghighipour}(2013)}]{kaltenegger2013b}
{Kaltenegger}, L., \& {Haghighipour}, N. 2013, \apj, 777, 165,
  \dodoi{10.1088/0004-637X/777/2/165}

\bibitem[{{Kane}(2023)}]{kane2023c}
{Kane}, S.~R. 2023, \aj, 166, 187, \dodoi{10.3847/1538-3881/acfb01}

\bibitem[{{Kane} \& {Burt}(2024)}]{kane2024e}
{Kane}, S.~R., \& {Burt}, J.~A. 2024, \aj, 168, 279,
  \dodoi{10.3847/1538-3881/ad8a68}

\bibitem[{{Kane} \& {Deveny}(2018)}]{kane2018b}
{Kane}, S.~R., \& {Deveny}, S.~J. 2018, \apj, 864, 115,
  \dodoi{10.3847/1538-4357/aad802}

\bibitem[{{Kane} \& {Gelino}(2012)}]{kane2012a}
{Kane}, S.~R., \& {Gelino}, D.~M. 2012, \pasp, 124, 323, \dodoi{10.1086/665271}

\bibitem[{{Kane} \& {Hinkel}(2013)}]{kane2013a}
{Kane}, S.~R., \& {Hinkel}, N.~R. 2013, \apj, 762, 7,
  \dodoi{10.1088/0004-637X/762/1/7}

\bibitem[{{Kane} {et~al.}(2024){Kane}, {Li}, {Turnbull}, {Dressing}, \&
  {Harada}}]{kane2024d}
{Kane}, S.~R., {Li}, Z., {Turnbull}, M.~C., {Dressing}, C.~D., \& {Harada},
  C.~K. 2024, \aj, 168, 195, \dodoi{10.3847/1538-3881/ad6a50}

\bibitem[{{Kane} {et~al.}(2020){Kane}, {Turnbull}, {Fulton}, {Rosenthal},
  {Howard}, {Isaacson}, {Marcy}, \& {Weiss}}]{kane2020b}
{Kane}, S.~R., {Turnbull}, M.~C., {Fulton}, B.~J., {et~al.} 2020, \aj, 160, 81,
  \dodoi{10.3847/1538-3881/ab9ffe}

\bibitem[{{Kane} {et~al.}(2016){Kane}, {Hill}, {Kasting}, {Kopparapu},
  {Quintana}, {Barclay}, {Batalha}, {Borucki}, {Ciardi}, {Haghighipour},
  {Hinkel}, {Kaltenegger}, {Selsis}, \& {Torres}}]{kane2016c}
{Kane}, S.~R., {Hill}, M.~L., {Kasting}, J.~F., {et~al.} 2016, \apj, 830, 1,
  \dodoi{10.3847/0004-637X/830/1/1}

\bibitem[{{Kane} {et~al.}(2022){Kane}, {Foley}, {Hill}, {Unterborn}, {Barclay},
  {Cale}, {Gilbert}, {Plavchan}, \& {Wittrock}}]{kane2022a}
{Kane}, S.~R., {Foley}, B.~J., {Hill}, M.~L., {et~al.} 2022, \aj, 163, 20,
  \dodoi{10.3847/1538-3881/ac366b}

\bibitem[{{Kasting} {et~al.}(1993){Kasting}, {Whitmire}, \&
  {Reynolds}}]{kasting1993a}
{Kasting}, J.~F., {Whitmire}, D.~P., \& {Reynolds}, R.~T. 1993, \icarus, 101,
  108, \dodoi{10.1006/icar.1993.1010}

\bibitem[{{Kipping}(2013)}]{kipping2013b}
{Kipping}, D.~M. 2013, \mnras, 434, L51, \dodoi{10.1093/mnrasl/slt075}

\bibitem[{{Kopparapu} \& {Barnes}(2010)}]{kopparapu2010}
{Kopparapu}, R.~K., \& {Barnes}, R. 2010, \apj, 716, 1336,
  \dodoi{10.1088/0004-637X/716/2/1336}

\bibitem[{{Kopparapu} {et~al.}(2014){Kopparapu}, {Ramirez}, {SchottelKotte},
  {Kasting}, {Domagal-Goldman}, \& {Eymet}}]{kopparapu2014}
{Kopparapu}, R.~K., {Ramirez}, R.~M., {SchottelKotte}, J., {et~al.} 2014, \apj,
  787, L29, \dodoi{10.1088/2041-8205/787/2/L29}

\bibitem[{{Kopparapu} {et~al.}(2013){Kopparapu}, {Ramirez}, {Kasting}, {Eymet},
  {Robinson}, {Mahadevan}, {Terrien}, {Domagal-Goldman}, {Meadows}, \&
  {Deshpande}}]{kopparapu2013a}
{Kopparapu}, R.~K., {Ramirez}, R., {Kasting}, J.~F., {et~al.} 2013, \apj, 765,
  131, \dodoi{10.1088/0004-637X/765/2/131}

\bibitem[{{Kraus} {et~al.}(2012){Kraus}, {Ireland}, {Hillenbrand}, \&
  {Martinache}}]{kraus2012b}
{Kraus}, A.~L., {Ireland}, M.~J., {Hillenbrand}, L.~A., \& {Martinache}, F.
  2012, \apj, 745, 19, \dodoi{10.1088/0004-637X/745/1/19}

\bibitem[{{Laliotis} {et~al.}(2023){Laliotis}, {Burt}, {Mamajek}, {Li},
  {Perdelwitz}, {Zhao}, {Butler}, {Holden}, {Rosenthal}, {Fulton}, {Feng},
  {Kane}, {Bailey}, {Carter}, {Crane}, {Furlan}, {Gnilka}, {Howell},
  {Laughlin}, {Shectman}, {Teske}, {Tinney}, {Vogt}, {Wang}, \&
  {Wittenmyer}}]{laliotis2023}
{Laliotis}, K., {Burt}, J.~A., {Mamajek}, E.~E., {et~al.} 2023, \aj, 165, 176,
  \dodoi{10.3847/1538-3881/acc067}

\bibitem[{{Malmberg} {et~al.}(2011){Malmberg}, {Davies}, \&
  {Heggie}}]{malmberg2011}
{Malmberg}, D., {Davies}, M.~B., \& {Heggie}, D.~C. 2011, \mnras, 411, 859,
  \dodoi{10.1111/j.1365-2966.2010.17730.x}

\bibitem[{{Malmberg} {et~al.}(2007){Malmberg}, {de Angeli}, {Davies}, {Church},
  {Mackey}, \& {Wilkinson}}]{malmberg2007b}
{Malmberg}, D., {de Angeli}, F., {Davies}, M.~B., {et~al.} 2007, \mnras, 378,
  1207, \dodoi{10.1111/j.1365-2966.2007.11885.x}

\bibitem[{{Matson} {et~al.}(2018){Matson}, {Howell}, {Horch}, \&
  {Everett}}]{matson2018}
{Matson}, R.~A., {Howell}, S.~B., {Horch}, E.~P., \& {Everett}, M.~E. 2018,
  \aj, 156, 31, \dodoi{10.3847/1538-3881/aac778}

\bibitem[{{Pearce} {et~al.}(2020){Pearce}, {Kraus}, {Dupuy}, {Mann}, {Newton},
  {Tofflemire}, \& {Vanderburg}}]{pearce2020}
{Pearce}, L.~A., {Kraus}, A.~L., {Dupuy}, T.~J., {et~al.} 2020, \apj, 894, 115,
  \dodoi{10.3847/1538-4357/ab8389}

\bibitem[{{Petigura} {et~al.}(2017){Petigura}, {Howard}, {Marcy}, {Johnson},
  {Isaacson}, {Cargile}, {Hebb}, {Fulton}, {Weiss}, {Morton}, {Winn}, {Rogers},
  {Sinukoff}, {Hirsch}, \& {Crossfield}}]{petigura2017a}
{Petigura}, E.~A., {Howard}, A.~W., {Marcy}, G.~W., {et~al.} 2017, \aj, 154,
  107, \dodoi{10.3847/1538-3881/aa80de}

\bibitem[{{Petrovich}(2015)}]{petrovich2015a}
{Petrovich}, C. 2015, \apj, 799, 27, \dodoi{10.1088/0004-637X/799/1/27}

\bibitem[{{Porto de Mello} {et~al.}(2006){Porto de Mello}, {del Peloso}, \&
  {Ghezzi}}]{portodemello2006}
{Porto de Mello}, G., {del Peloso}, E.~F., \& {Ghezzi}, L. 2006, Astrobiology,
  6, 308, \dodoi{10.1089/ast.2006.6.308}

\bibitem[{{Quarles} {et~al.}(2022){Quarles}, {Li}, \& {Lissauer}}]{quarles2022}
{Quarles}, B., {Li}, G., \& {Lissauer}, J.~J. 2022, \mnras, 509, 2736,
  \dodoi{10.1093/mnras/stab3179}

\bibitem[{{Quarles} \& {Lissauer}(2016)}]{quarles2016}
{Quarles}, B., \& {Lissauer}, J.~J. 2016, \aj, 151, 111,
  \dodoi{10.3847/0004-6256/151/5/111}

\bibitem[{{Rabl} \& {Dvorak}(1988)}]{rabl1988}
{Rabl}, G., \& {Dvorak}, R. 1988, \aap, 191, 385

\bibitem[{{Radovan} {et~al.}(2014){Radovan}, {Lanclos}, {Holden}, {Kibrick},
  {Allen}, {Deich}, {Rivera}, {Burt}, {Fulton}, {Butler}, \&
  {Vogt}}]{radovan2014}
{Radovan}, M.~V., {Lanclos}, K., {Holden}, B.~P., {et~al.} 2014, Society of
  Photo-Optical Instrumentation Engineers (SPIE) Conference Series, Vol. 9145,
  {The automated planet finder at Lick Observatory} (SPIE Press), 91452B,
  \dodoi{10.1117/12.2057310}

\bibitem[{{Raymond} {et~al.}(2024){Raymond}, {Kaib}, {Selsis}, \&
  {Bouy}}]{raymond2024a}
{Raymond}, S.~N., {Kaib}, N.~A., {Selsis}, F., \& {Bouy}, H. 2024, \mnras, 527,
  6126, \dodoi{10.1093/mnras/stad3604}

\bibitem[{{Rein} \& {Liu}(2012)}]{rein2012a}
{Rein}, H., \& {Liu}, S.~F. 2012, \aap, 537, A128,
  \dodoi{10.1051/0004-6361/201118085}

\bibitem[{{Rein} \& {Tamayo}(2015)}]{rein2015c}
{Rein}, H., \& {Tamayo}, D. 2015, \mnras, 452, 376,
  \dodoi{10.1093/mnras/stv1257}

\bibitem[{{Robertson} {et~al.}(2019){Robertson}, {Anderson}, {Stefansson},
  {Hearty}, {Monson}, {Mahadevan}, {Blakeslee}, {Bender}, {Ninan}, {Conran},
  {Levi}, {Lubar}, {Cole}, {Dykhouse}, {Kanodia}, {Nitroy}, {Smolsky},
  {Tuggle}, {Blank}, {Nelson}, {Blake}, {Halverson}, {Henderson}, {Kaplan},
  {Li}, {Logsdon}, {McElwain}, {Rajagopal}, {Ramsey}, {Roy}, {Schwab},
  {Terrien}, \& {Wright}}]{robertson2019}
{Robertson}, P., {Anderson}, T., {Stefansson}, G., {et~al.} 2019, Journal of
  Astronomical Telescopes, Instruments, and Systems, 5, 015003,
  \dodoi{10.1117/1.JATIS.5.1.015003}

\bibitem[{{Rosenthal} {et~al.}(2021){Rosenthal}, {Fulton}, {Hirsch},
  {Isaacson}, {Howard}, {Dedrick}, {Sherstyuk}, {Blunt}, {Petigura}, {Knutson},
  {Behmard}, {Chontos}, {Crepp}, {Crossfield}, {Dalba}, {Fischer}, {Henry},
  {Kane}, {Kosiarek}, {Marcy}, {Rubenzahl}, {Weiss}, \&
  {Wright}}]{rosenthal2021}
{Rosenthal}, L.~J., {Fulton}, B.~J., {Hirsch}, L.~A., {et~al.} 2021, \apjs,
  255, 8, \dodoi{10.3847/1538-4365/abe23c}

\bibitem[{{Schwab} {et~al.}(2016){Schwab}, {Rakich}, {Gong}, {Mahadevan},
  {Halverson}, {Roy}, {Terrien}, {Robertson}, {Hearty}, {Levi}, {Monson},
  {Wright}, {McElwain}, {Bender}, {Blake}, {St{\"u}rmer}, {Gurevich},
  {Chakraborty}, \& {Ramsey}}]{schwab2016}
{Schwab}, C., {Rakich}, A., {Gong}, Q., {et~al.} 2016, in Society of
  Photo-Optical Instrumentation Engineers (SPIE) Conference Series, Vol. 9908,
  Ground-based and Airborne Instrumentation for Astronomy VI, ed. C.~J.
  {Evans}, L.~{Simard}, \& H.~{Takami}, 99087H, \dodoi{10.1117/12.2234411}

\bibitem[{{Silva Aguirre} {et~al.}(2015){Silva Aguirre}, {Davies}, {Basu},
  {Christensen-Dalsgaard}, {Creevey}, {Metcalfe}, {Bedding}, {Casagrande},
  {Handberg}, {Lund}, {Nissen}, {Chaplin}, {Huber}, {Serenelli}, {Stello}, {Van
  Eylen}, {Campante}, {Elsworth}, {Gilliland}, {Hekker}, {Karoff}, {Kawaler},
  {Kjeldsen}, \& {Lundkvist}}]{silvaaguirre2015}
{Silva Aguirre}, V., {Davies}, G.~R., {Basu}, S., {et~al.} 2015, \mnras, 452,
  2127, \dodoi{10.1093/mnras/stv1388}

\bibitem[{{Strand}(1969)}]{strand1969}
{Strand}, K.~A. 1969, \aj, 74, 760, \dodoi{10.1086/110853}

\bibitem[{{Takeda} {et~al.}(2008){Takeda}, {Kita}, \& {Rasio}}]{takeda2008}
{Takeda}, G., {Kita}, R., \& {Rasio}, F.~A. 2008, \apj, 683, 1063,
  \dodoi{10.1086/589852}

\bibitem[{{Touma} \& {Sridhar}(2015)}]{touma2015}
{Touma}, J.~R., \& {Sridhar}, S. 2015, \nat, 524, 439,
  \dodoi{10.1038/nature14873}

\bibitem[{{Tuchow} {et~al.}(2024){Tuchow}, {Stark}, \& {Mamajek}}]{tuchow2024}
{Tuchow}, N.~W., {Stark}, C.~C., \& {Mamajek}, E. 2024, \aj, 167, 139,
  \dodoi{10.3847/1538-3881/ad25ec}

\bibitem[{{Turnbull}(2015)}]{turnbull2015}
{Turnbull}, M.~C. 2015, arXiv e-prints, arXiv:1510.01731.
\newblock \doarXiv{1510.01731}

\bibitem[{{van de Kamp}(1948)}]{vanderkamp1948}
{van de Kamp}, P. 1948, \aj, 53, 165, \dodoi{10.1086/106087}

\bibitem[{{Veras} \& {Raymond}(2012)}]{veras2012b}
{Veras}, D., \& {Raymond}, S.~N. 2012, \mnras, 421, L117,
  \dodoi{10.1111/j.1745-3933.2012.01218.x}

\bibitem[{{Vogt}(1987)}]{vogt1987b}
{Vogt}, S.~S. 1987, \pasp, 99, 1214, \dodoi{10.1086/132107}

\bibitem[{{Vogt} {et~al.}(1994){Vogt}, {Allen}, {Bigelow}, {Bresee}, {Brown},
  {Cantrall}, {Conrad}, {Couture}, {Delaney}, {Epps}, {Hilyard}, {Hilyard},
  {Horn}, {Jern}, {Kanto}, {Keane}, {Kibrick}, {Lewis}, {Osborne},
  {Pardeilhan}, {Pfister}, {Ricketts}, {Robinson}, {Stover}, {Tucker}, {Ward},
  \& {Wei}}]{vogt1994}
{Vogt}, S.~S., {Allen}, S.~L., {Bigelow}, B.~C., {et~al.} 1994, Society of
  Photo-Optical Instrumentation Engineers (SPIE) Conference Series, Vol. 2198,
  {HIRES: the high-resolution echelle spectrometer on the Keck 10-m Telescope}
  (SPIE Press), 362, \dodoi{10.1117/12.176725}

\bibitem[{{Vogt} {et~al.}(2014){Vogt}, {Radovan}, {Kibrick}, {Butler},
  {Alcott}, {Allen}, {Arriagada}, {Bolte}, {Burt}, {Cabak}, {Chloros},
  {Cowley}, {Deich}, {Dupraw}, {Earthman}, {Epps}, {Faber}, {Fischer}, {Gates},
  {Hilyard}, {Holden}, {Johnston}, {Keiser}, {Kanto}, {Katsuki}, {Laiterman},
  {Lanclos}, {Laughlin}, {Lewis}, {Lockwood}, {Lynam}, {Marcy}, {McLean},
  {Miller}, {Misch}, {Peck}, {Pfister}, {Phillips}, {Rivera}, {Sand ford},
  {Saylor}, {Stover}, {Thompson}, {Walp}, {Ward}, {Wareham}, {Wei}, \&
  {Wright}}]{vogt2014a}
{Vogt}, S.~S., {Radovan}, M., {Kibrick}, R., {et~al.} 2014, \pasp, 126, 359,
  \dodoi{10.1086/676120}

\bibitem[{{von Braun} {et~al.}(2014){von Braun}, {Boyajian}, {van Belle},
  {Kane}, {Jones}, {Farrington}, {Schaefer}, {Vargas}, {Scott}, {ten
  Brummelaar}, {Kephart}, {Gies}, {Ciardi}, {L{\'o}pez-Morales}, {Mazingue},
  {McAlister}, {Ridgway}, {Goldfinger}, {Turner}, \& {Sturmann}}]{vonbraun2014}
{von Braun}, K., {Boyajian}, T.~S., {van Belle}, G.~T., {et~al.} 2014, \mnras,
  438, 2413, \dodoi{10.1093/mnras/stt2360}

\bibitem[{{Wang} {et~al.}(2014){Wang}, {Fischer}, {Xie}, \&
  {Ciardi}}]{wang2014b}
{Wang}, J., {Fischer}, D.~A., {Xie}, J.-W., \& {Ciardi}, D.~R. 2014, \apj, 791,
  111, \dodoi{10.1088/0004-637X/791/2/111}

\bibitem[{{Wang} \& {Cuntz}(2017)}]{wang2017d}
{Wang}, Z., \& {Cuntz}, M. 2017, \aj, 154, 157,
  \dodoi{10.3847/1538-3881/aa8621}

\bibitem[{{Wittenmyer} {et~al.}(2020){Wittenmyer}, {Clark}, {Sharma}, {Stello},
  {Horner}, {Kane}, {Stevens}, {Wright}, {Spina}, {{\v{C}}otar}, {Asplund},
  {Bland-Hawthorn}, {Buder}, {Casey}, {De Silva}, {D'Orazi}, {Freeman}, {Kos},
  {Lewis}, {Lin}, {Lind}, {Martell}, {Simpson}, {Zucker}, \&
  {Zwitter}}]{wittenmyer2020c}
{Wittenmyer}, R.~A., {Clark}, J.~T., {Sharma}, S., {et~al.} 2020, \mnras, 496,
  851, \dodoi{10.1093/mnras/staa1528}

\bibitem[{{Wittrock} {et~al.}(2016){Wittrock}, {Kane}, {Horch}, {Hirsch},
  {Howell}, {Ciardi}, {Everett}, \& {Teske}}]{wittrock2016}
{Wittrock}, J.~M., {Kane}, S.~R., {Horch}, E.~P., {et~al.} 2016, \aj, 152, 149,
  \dodoi{10.3847/0004-6256/152/5/149}

\bibitem[{{Wittrock} {et~al.}(2017){Wittrock}, {Kane}, {Horch}, {Howell},
  {Ciardi}, \& {Everett}}]{wittrock2017}
---. 2017, \aj, 154, 184, \dodoi{10.3847/1538-3881/aa8d69}

\bibitem[{{Ziegler} {et~al.}(2020){Ziegler}, {Tokovinin}, {Brice{\~n}o},
  {Mang}, {Law}, \& {Mann}}]{ziegler2020}
{Ziegler}, C., {Tokovinin}, A., {Brice{\~n}o}, C., {et~al.} 2020, \aj, 159, 19,
  \dodoi{10.3847/1538-3881/ab55e9}

\bibitem[{{Zink} {et~al.}(2020){Zink}, {Batygin}, \& {Adams}}]{zink2020c}
{Zink}, J.~K., {Batygin}, K., \& {Adams}, F.~C. 2020, \aj, 160, 232,
  \dodoi{10.3847/1538-3881/abb8de}

\end{thebibliography}


\end{document}